\documentclass [12pt]{dafne}

\begin{document}

\title{
\vspace{-1.5cm}
\rightline{\small\rm BNL--67140}
\vspace{-0.3cm}
\rightline{\small\rm February 3, 2000}
\vspace{1cm}
Experimental Results on Radiative Kaon Decays\footnote{
To be published in the Proceedings of the 3${rd}$ Workshop on
PHYSICS AND DETECTORS FOR DA$\Phi$NE, Frascati, Italy, Nov. 16--19, 1999 
   (Frascati Physics Series Vol.~XVI, 2000)}
}
\author{
S.~Kettell                            \\
{\em Brookhaven National Laboratory} \\
}
\date{}
\maketitle
\begin{abstract}
This paper reviews the current status of experimental results
on radiative kaon decays. Several experiments at BNL, CERN and FNAL
have recently or will soon complete data collection; as a
result, there are several new results.
\end{abstract}

\baselineskip=17pt

\section{Introduction}

Radiative kaon decays provide a testing ground for Chiral Perturbation
Theory (ChPT). ChPT provides a framework for calculating the decay
rates for several modes, either directly or relative to other measured
modes. The radiative modes are important for determining long
distance contributions to other decays of interest: the two-photon
contribution to $K^\circ_L \! \rightarrow \! \mu^+ \mu^-$, and the
CP-conserving and indirect CP-violating contributions to $K^\circ_L \!
\rightarrow \! \pi^\circ e^+ e^-$ and $K^\circ_L \! \rightarrow \!
\pi^\circ \mu^+ \mu^-$.  They are also important as backgrounds to
other modes (e.g. the $K^\circ_L \! \rightarrow \! e^+
e^-\gamma\gamma$ background to $K^\circ_L \! \rightarrow \! \pi^\circ
e^+ e^-$).

A number of recent results have been reported in the literature, as
well as in several recent
conferences\cite{ichep98,dpf99,moriond99,panic99,epshep99,kaon99}.

\section{Radiative K$_{\pi2}$ Decays}

The radiative K$_{\pi2}$ decays: $K^+ \! \rightarrow \! \pi^+
\pi^\circ \gamma$, $K^\circ_L \! \rightarrow \! \pi^+ \pi^- \gamma$
and $K^\circ_{\rm S} \! \rightarrow \! \pi^+ \pi^- \gamma$ have two
contributions.  One is inner bremsstrahlung (IB) radiation from one of
the charged particles. The second is direct emission (DE) from the
vertex.  The branching ratio of the IB contribution scales with the
underlying K$_{\pi2}$ decay rate. Whereas, the rate for direct
emission is expected to be roughly comparable for all three modes.

\begin{figure}[ht]
\vspace{3.5cm}
 \includegraphics{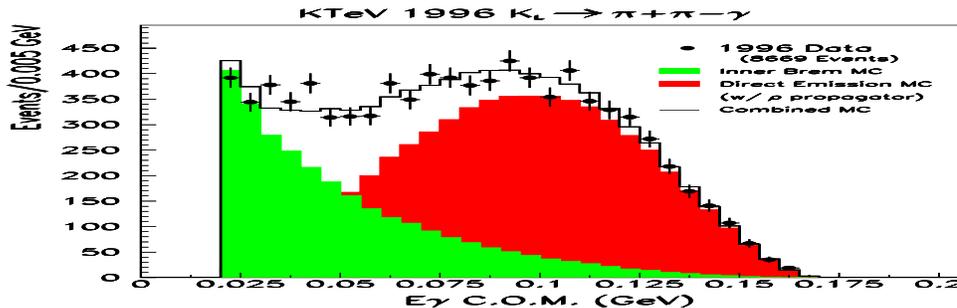}
 \caption{\it Photon energy with fit to IB and DE (with $\rho$
 propagator) from KTeV. \label{fig_klppg} }
\end{figure}

A new result\cite{klppg} for $K^\circ_L \! \rightarrow \! \pi^+ \pi^-
\gamma$ from KTeV is shown in Fig.~\ref{fig_klppg}.  The energy of
the photon is shown, along with the contributions from IB and DE.
The DE component is modified by a ``$\rho$-propagator'' that serves to
soften the DE spectrum.  The branching ratio for the direct emission
component (see eq.\ref{eqn_klppg}) is
\begin{equation}
{\rm BR(K^\circ_L \! \rightarrow \! \pi^+ \pi^- \gamma;DE)=
(3.70\pm0.10)\times 10^{-5} (E_{\gamma}^{*}>20 MeV) }
\label{eqn_klppg}
\end{equation}
The ratio of direct emission to DE+IB is (see eq.\ref{eqn_klppg_de})
\begin{equation}
{\rm DE/(DE+IB) = 0.685\pm 0.009\pm0.017 }
\label{eqn_klppg_de}
\end{equation}
This result is based on $\sim$5\% of the total KTeV data for this
mode.

\begin{figure}[ht]
 \begin{minipage}{0.3\linewidth}
  \includegraphics{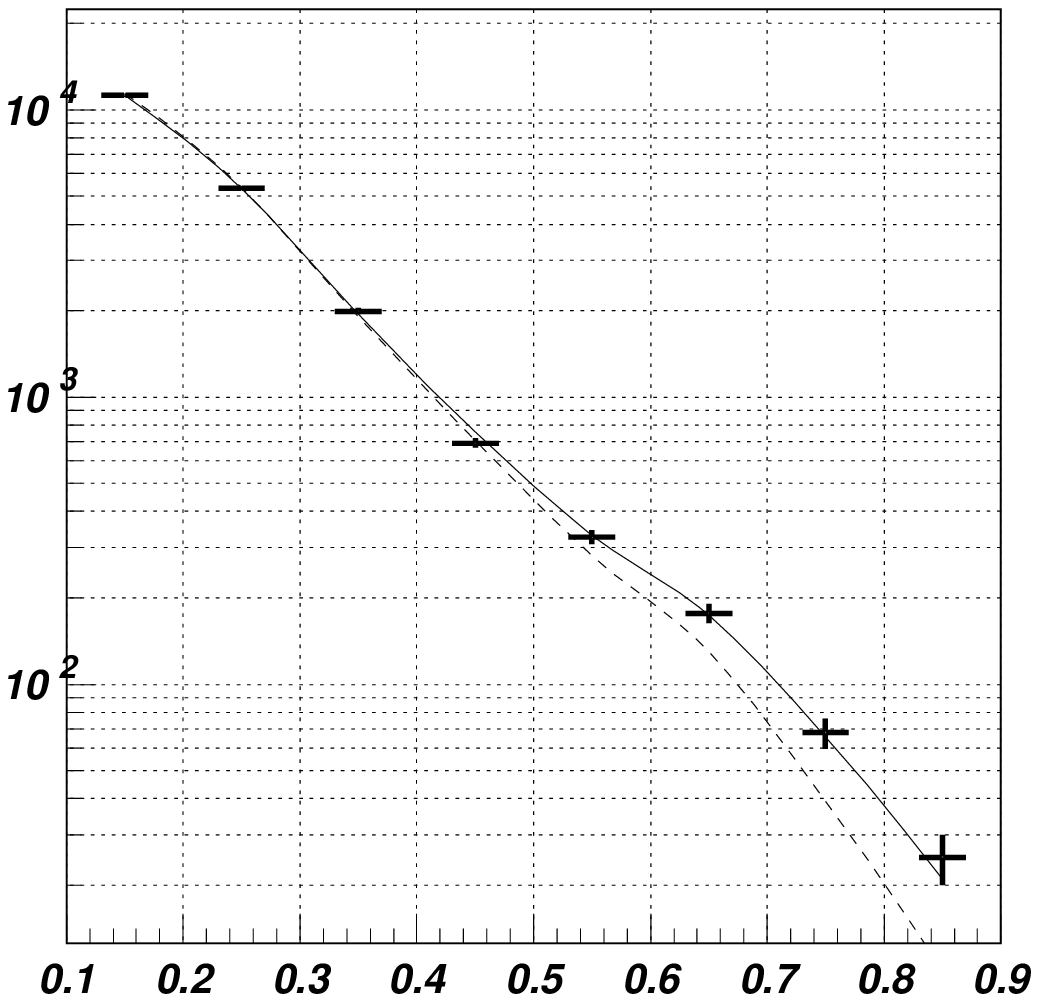}
 \end{minipage}\hfill
 \begin{minipage}{0.3\linewidth}
  \includegraphics{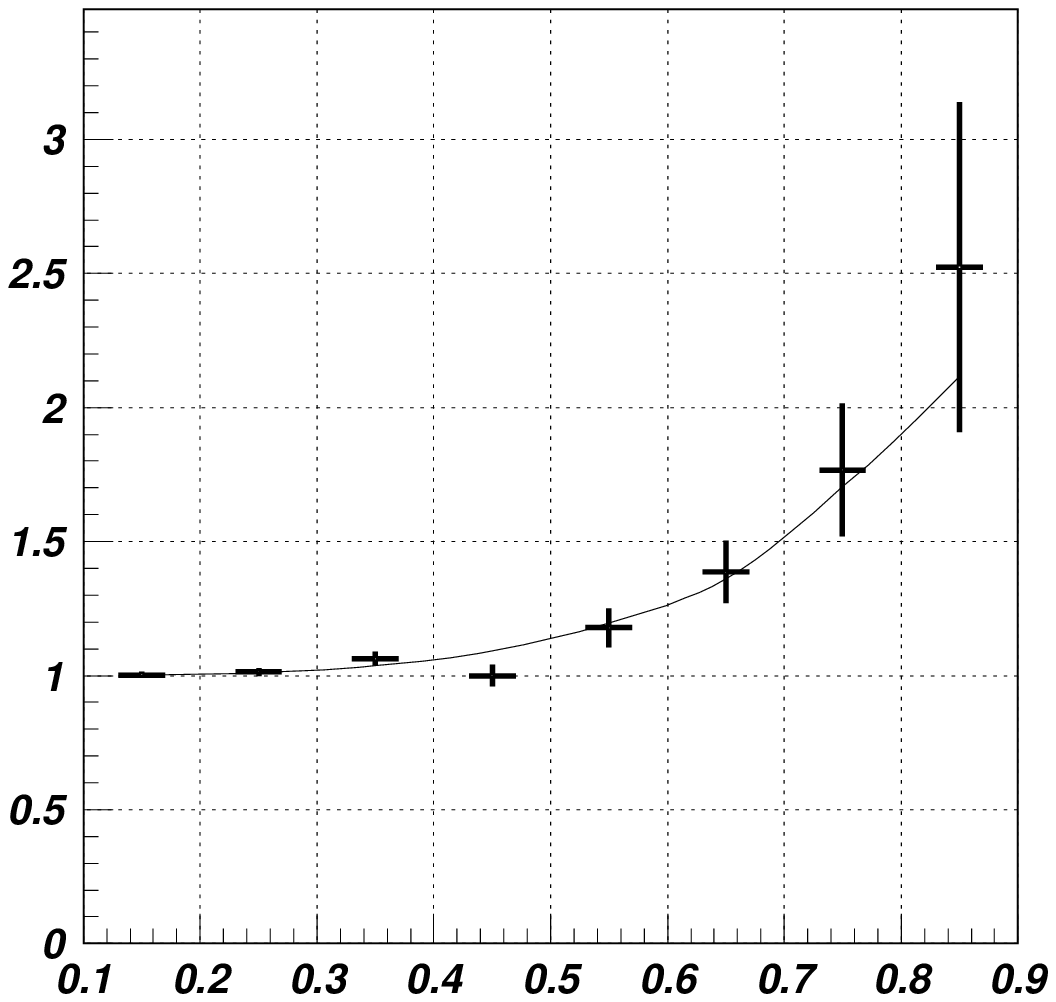}
 \end{minipage}\hfill
 \caption{\it (left) W spectrum of the signal events and best fits to
 IB+DE [solid curve] and IB alone [dashed curve]; (right) W spectrum
 normalized to the IB spectrum.  \label{fig_kppg} }
\end{figure}

There are new results from E787\cite{k+ppg} in the charged decay mode
($K^+ \! \rightarrow \! \pi^+ \pi^\circ \gamma$) as well.  This result
is striking, in that the branching ratio is a factor of 4 lower than
the previous value.  The data is traditionally expressed in terms of
the variable W, which is defined as:
\begin{eqnarray}
{\rm W^{2}} & \equiv & {\rm (p\cdot q)/{m_{K^+}^{2}} \times
(p_{+}\cdot q)/{m_{\pi^+}^{2}}} \\ \nonumber
           &   =  & {\rm E_{\gamma}^2\times (E_{\pi^+} - P_{\pi^+}\times
\cos{\bf\theta_{\pi^+\;\gamma}})/({m_{K^+}^2}\times{m_{\pi^+}^{2}})}
\label{eqn_w}
\end{eqnarray}
The new result from E787, shown in Figure~\ref{fig_kppg},
\vspace{3.5cm}
has about 8 times higher statistics than the old one.  The branching
ratio for the direct emission component, from a fit to IB and DE (see
eq.\ref{eqn_kppg}) is
\begin{equation}
{\rm BR(K^+ \! \rightarrow \! \pi^+ \pi^\circ \gamma;DE)=
(4.72\pm0.77)\times 10^{-6} \, (55<\!T_{\pi^+}\!<90 MeV) }
\label{eqn_kppg}
\end{equation}
This represents half of the E787 data that is currently on tape.  The
interference term is small, $(-0.4\pm1.6)$\% and the direct emission
is $(1.85\pm0.30)$\%. The decay rate, corrected to full phase
space\footnote{This correction assumes that the form factor has no
energy dependence.}, is now measured to be similar to that for $K_L$:
$\Gamma(K^+ \! \rightarrow \! \pi^+ \pi^\circ \gamma;DE)=808\pm132
s^{-1}$ vs. $\Gamma(K^\circ_L \! \rightarrow \! \pi^+ \pi^-
\gamma;DE)=617\pm18 s^{-1}$.

KTeV also has new results on $K^\circ_L \! \rightarrow \! \pi^+ \pi^-
e^+ e^- $, where the photon has internally converted to two
electrons. In addition to measuring the branching ratio\cite{klppee},
a T-odd observable in the angular distribution of the plane of the
$\pi$-pair vs. the plane of the electron pair is
observed\cite{klppee2}.  This data represents one quarter of the
final KTeV sample.

A summary of the current experimental status of radiative K$_{\pi2}$
decays is shown in Table~\ref{tab_kzppg}.
\begin{table}[ht]
\centering
\caption{ \it Summary of Radiative K$_{\pi2}$ results. }
\vskip 0.1 in
\begin{tabular}{|l||l|l|} \hline
Decay Mode  &  \multicolumn{1}{|c|}{Branching Ratio} & Citation \\
 \hline\hline
$K^\circ_L \! \rightarrow \! \pi^+\pi^-\gamma(DE)$
& $(3.70\pm0.10)\times10^{-5}$ & KTeV-99\cite{klppg}\\ \hline
$K^+ \! \rightarrow \! \pi^+\pi^\circ\gamma(DE)$
&$(4.72\pm0.77)\times10^{-6}$ & E787-99\cite{k+ppg}\\ \hline
$K^\circ_L \! \rightarrow \! \pi^+\pi^-e^+e^-$  &
$(3.63\pm.11\pm.14)\times10^{-7}$ & KTeV-99\cite{klppee}\\ \hline
$K^\circ_L \! \rightarrow \! \pi^\circ\pi^\circ\gamma$  &
$< 5.6\times10^{-6}$ & NA31-94\cite{klp0p0g}\\ \hline
$K^\circ_S \! \rightarrow \! \pi^+\pi^-\gamma$  &
$(1.78\pm0.05)\times10^{-3}$ & E731-93\cite{ksppg}\\
$K^\circ_S \! \rightarrow \! \pi^+\pi^-\gamma(DE)$  &
$<0.06\times10^{-3}$ & CERN-76\cite{ksppg_de}\\ \hline
\end{tabular}
\label{tab_kzppg}
\end{table}

\section{$K \! \rightarrow \! \pi \gamma \gamma$ Decays}

The decay $K^\circ_L \! \rightarrow \! \pi^\circ \gamma\gamma$ is very
interesting, since to ${\cal O}(p^4)$ of ChPT the decay rate and
spectral shape are completely determined, without any free
parameters\cite{chpt_klpgg}. The prediction of the spectral shape is a
striking success of ChPT; however, the decay rate is a factor of 3 too
small. To match the experimental number a model dependent contribution
from ${\cal O}(p^6)$ is needed, which is usually parameterized with a
constant $a_V$\cite{chpt_av}.  The CP-conserving contribution to
$K^\circ_L \! \rightarrow \! \pi^\circ e^+ e^-$ depends on the value
of $a_V$.  Based on half of the total data sample, KTeV has recently
measured $a_V = -0.72\pm0.05\pm0.06$\cite{klpgg}, implying a
contribution of $1-2\times10^{-12}$.

The charged mode $K^+ \! \rightarrow \! \pi^+ \gamma \gamma$ is more
complicated, requiring an unknown parameter, $\hat{c}$, even at ${\cal
O}(p^4)$. Both the decay rate and spectral shape are predicted with
this single parameter. E787 has measured $\hat{c} =
1.8\pm0.6$\cite{kpgg}.

The experimental measurements of $K \! \rightarrow \! \pi \gamma
\gamma$ are summarized in Table~\ref{tab_kzpgg}.
\begin{table}[ht]
\centering
\caption{ \it Summary of $K \! \rightarrow \! \pi \gamma \gamma$ results. }
\vskip 0.1 in
\begin{tabular}{|l||l|l|} \hline
Decay Mode  &  \multicolumn{1}{|c|}{Branching Ratio} & Citation \\ \hline\hline
$K^\circ_L \! \rightarrow \! \pi^\circ\gamma\gamma$
 & $(1.68\pm0.07\pm0.08)\times10^{-6}$ & KTeV-99\cite{klpgg}\\ \hline
$K^+ \! \rightarrow \! \pi^+\gamma\gamma$ &
 $(6.0\pm1.5\pm0.7)\times10^{-7}$& E787-97\cite{kpgg}\\ \hline
 $K^\circ_S \! \rightarrow \! \pi^\circ\gamma\gamma$  &
no limit & (NA48-02?)\\ \hline
 $K^\circ_L \! \rightarrow \! \pi^\circ e^+e^-$  &
$< 5.6\times10^{-10}$ & KTeV-99\cite{klpee}\\ \hline
 $K^\circ_L \! \rightarrow \! \pi^\circ\mu^+\mu^-$  &
$< 3.4\times10^{-10}$ & KTeV-99\cite{klpmm}\\ \hline
$K^+ \! \rightarrow \! \pi^+e^+e^-$  &
$(2.94\pm0.05\pm0.13)\times10^{-7}$ & E865-99\cite{kpee}\\ \hline
$K^+ \! \rightarrow \! \pi^+\mu^+\mu^-$ &
$(9.22\pm0.60\pm0.49)\times10^{-8}$ & E865-99\cite{kpmm}\\ \hline
$K^\circ_S \! \rightarrow \! \pi^\circ e^+e^-$ &
$< 1.1\times10^{-6}$ & NA31-93\cite{kspee}\\ \hline
$K^\circ_L \! \rightarrow \! \pi^\circ e^+e^-\gamma$ &
$(2.20\pm0.48\pm0.11)\times10^{-8}$ & KTeV-99\cite{klpeeg}\\ \hline
$K^+ \! \rightarrow \! \pi^+e^+e^-\gamma$ & $\sim$30 events &
E865-99\cite{kpeeg}\\ \hline
\end{tabular}
\label{tab_kzpgg}
\end{table}
The KTeV measurement of $K^\circ_L \! \rightarrow \! \pi^\circ e^+ e^-
\gamma $ should improve by $\times$3; the measurements of $K^\circ_L
\! \rightarrow \! \pi^\circ \ell^+ \ell^-$ are background limited,
and will improve by $\sqrt{3}$.

\section{K$^0$ to Two Real or Off-shell Photons}

The decay $K^\circ_{\rm S} \! \rightarrow \! \gamma \gamma$ is
predicted in ${\cal O}(p^4)$ of ChPT, without any free parameters, to
occur with BR($K^\circ_{\rm S} \! \rightarrow \! \gamma \gamma$) =
$2.0\times10^{-6}$\cite{chpt_klpgg}.  This is in good agreement with
the experimental value\cite{ksgg} (see Table~\ref{tab_kgg}), although
the experimental errors need to be reduced.

The decay $K^\circ_L \! \rightarrow \!  \gamma\gamma$ is of interest
for its importance in interpreting the measurement of $K^\circ_L \!
\rightarrow \! \mu^+ \mu^-$.  The decay $K^\circ_L \! \rightarrow \!
\mu^+ \mu^-$ is sensitive to internal top quark loops, that would
allow a determination of the fundamental SM parameter $\rho$.  The
decay is, however, dominated by the decay $K^\circ_L \! \rightarrow \!
\gamma\gamma$ with the photons converting to a $\mu^\pm$ pair. For
this reason a precise measure of $K^\circ_L \! \rightarrow \!
\gamma\gamma$ is needed. With the improved precision on $K^\circ_L \!
\rightarrow \! \mu^+ \mu^-$ from E871, the uncertainties on $K^\circ_L
\! \rightarrow \!  \gamma\gamma$ and $\frac{K^\circ_{\rm L} \!
\rightarrow \! \pi^\circ \pi^\circ} {K^\circ_{\rm S} \! \rightarrow \!
\pi^+ \pi^- }$ are now contributing significantly\cite{klgg,klmm,pdg}
to the uncertainty on the ratio
\begin{eqnarray}
\frac{\Gamma(K^\circ_L \! \rightarrow \! \mu^+ \mu^-)}
{\Gamma(K^\circ_L \! \rightarrow \!  \gamma\gamma)} & = &
\left[ \begin{array}{c}B(K^\circ_L \! \rightarrow \! \mu^+ \mu^-) \\
\hline B(K^\circ_L \! \rightarrow \! \pi^+ \pi^-)\end{array} \right] \times \\
\nonumber
& & \hspace{-1.5cm} \left[ \left| \begin{array}{c}\eta_{+-}\\ \hline
\eta_{\circ\circ} \end{array} \right|
 \begin{array}{c}B(K^\circ_S \! \rightarrow \! \pi^+ \pi^-)
\\ \hline B(K^\circ_{\rm S} \! \rightarrow \! \pi^\circ \pi^\circ) \end{array}
\right]  \times
\left[  \begin{array}{c}B(K^\circ_{\rm L} \! \rightarrow \! \pi^\circ
\pi^\circ)
\\ \hline B(K^\circ_L \! \rightarrow \!  \gamma\gamma)
\end{array} \right] \\ \nonumber
& & [1.55\%] [(0.23\%)(1.28\%)][1.42\%] \\ \nonumber
& = & (1.213\pm0.030)\times10^{-5} \nonumber
\label{eqn_kmm}
\end{eqnarray}
KLOE should be able to contribute to improving both of these
measurements.  Finally there is a long distance dispersive
contribution, from two off-shell photons, for which additional input
from ChPT and measurements of the decays $K^\circ_L \! \rightarrow \!
e^+ e^-\gamma$, $K^\circ_L \! \rightarrow \! \mu^+ \mu^- e^+ e^-$ and
$K^\circ_L \! \rightarrow \! e^+ e^- e^+ e^-$ are
needed\cite{chpt_klmm}.

Results of kaon decays to two real or off-shell photons are summarized
in Table~\ref{tab_kgg}.
\begin{table}[ht]
\centering
\caption{ \it Summary of results of decays to two photons. }
\vskip 0.1 in
\begin{tabular}{|l||l|l|} \hline
Decay Mode  &  \multicolumn{1}{|c|}{Branching Ratio} & Citation \\ \hline\hline
$K^\circ_S \! \rightarrow \! \gamma\gamma$ & $(2.4\pm0.9)\times10^{-6}$ &
NA31-95\cite{ksgg}\\ \hline
$K^\circ_L \! \rightarrow \! \gamma\gamma$ & $(5.92\pm0.15)\times10^{-4}$ &
NA31-87\cite{klgg}\\ \hline
$K^\circ_L \! \rightarrow \!  \mu^+\mu^-$ & $(7.24\pm0.17)\times10^{-9}$ &
E871-99\cite{klmm}\\ \hline
$K^\circ_L \! \rightarrow \!  e^+e^-$ & $8.7^{+5.7}_{-4.1}\times10^{-12}$ &
E871-98\cite{klee}\\ \hline
$K^\circ_L\rightarrow e^+e^-\gamma$
&$\!(1.06\pm\!.02\pm\!.02\pm\!.04)\!\times\!10^{-5}\!$& NA48-99\cite{kleeg}\\
\hline
$K^\circ_L \! \rightarrow \!  \mu^+\mu^-\gamma$ &
$(3.23\pm0.23\pm0.19)\times10^{-7}$ & E799-95\cite{klmmg} \\ \hline
$K^\circ_L \! \rightarrow \!  e^+e^-e^+e^-$ &
$(4.14\pm0.27\pm0.31)\times10^{-8}$ & KTeV-98\cite{kleeee}\\ \hline
$K^\circ_L \! \rightarrow \!  e^+e^-\mu^+\mu^-$ & $\sim$40 events &
KTeV-99\cite{klmmee}\\ \hline
$K^\circ_L \! \rightarrow \!  \! \mu^+\mu^-\mu^+\mu^-$ & no limit & \\ \hline
$K^\circ_S \! \rightarrow \!  \mu^+\mu^-$ & $< 3.2\times10^{-7}$ &
CERN-73\cite{ksmm}\\ \hline
$K^\circ_S \! \rightarrow \!  e^+e^-$ & $< 1.4\times10^{-7}$ &
CPLEAR\cite{ksee}\\ \hline
$K^\circ_L \! \rightarrow \!  e^+e^-\gamma\gamma$ &
$(6.31\pm0.14\pm0.42)\times10^{-7}$ & KTeV-99\cite{kleegg}\\ \hline
$K^\circ_L \! \rightarrow \!  \mu^+\mu^-\gamma\gamma$ &
$(1.42^{+1.02}_{-0.81}\pm0.14)\times10^{-9}$ & KTeV-99\cite{klmmgg}\\ \hline
\end{tabular}
\label{tab_kgg}
\end{table}
The KTeV measurements of $K^\circ_L \! \rightarrow \! e^+ e^- e^+ e^-$
and $K^\circ_L \! \rightarrow \! \mu^+ \mu^- e^+ e^-$ should improve
by $\times$4 and the modes $K^\circ_L \! \rightarrow \! e^+
e^-\gamma\gamma$ and $K^\circ_L \! \rightarrow \! \mu^+ \mu^-
\gamma\gamma$ should improve by $\times$3 with the final KTeV data
set.  The $K^\circ_L \! \rightarrow \! e^+ e^-\gamma$ should improve
by $\times$20 and $K^\circ_L \! \rightarrow \! \mu^+ \mu^- \gamma$
should improve by $\times$30.  The $K_S$ modes may be improved by NA48
in a special run, after $\epsilon'/\epsilon$.  The $K^\circ_L \!
\rightarrow \!  \gamma\gamma$ and $K^\circ_S \! \rightarrow \!
\gamma\gamma$ as well as several other modes will be improved by KLOE.
There is no improvement in the foreseeable future for $K^\circ_L \!
\rightarrow \! e^+ e^-$ or $K^\circ_L \! \rightarrow \! \mu^+ \mu^-$.

\section{Radiative K$_{\ell2}$ Decays}

The form factors in the decays $K^+ \! \rightarrow \! \ell^+\nu_\ell
\gamma$, A and V, and, R, in the decays $K^+ \! \rightarrow \!
\ell^+\nu_\ell \ell'^+\ell'^-$, are predicted by ChPT. Recent
measurements should allow precise experimental determinations of all
three parameters.  The most recent determination of $|F_V + F_A| =
0.165\pm0.007\pm0.011$ from the E787 measurement of the direct
emission component of $K^+ \! \rightarrow \! \mu^+\nu\gamma$, usually
called Structure Dependent (SD$^+$) radiation, is consistent with the
previous determination of $|F_V + F_A| = 0.148\pm0.010$ from $K^+ \!
\rightarrow \!  e^+\nu\gamma$.  A limit of $-0.25 < F_V - F_A < 0.07$
is derived from the $K^+ \! \rightarrow \! \mu^+\nu\gamma(SD^+)$.  An
improved measure of $F_V - F_A$ along with a measure of $R$ should be
available soon from E865.

A summary of the recent radiative K$_{\ell2}$ results is presented in
Table~\ref{tab_kl2g}.
\begin{table}[ht]
\centering
\caption{ \it Summary of Radiative K$_{\ell2}$ results. }
\vskip 0.1 in
\begin{tabular}{|l||l|l|} \hline
Decay Mode  &  \multicolumn{1}{|c|}{Branching Ratio} & Citation \\ \hline\hline
$K^+ \! \rightarrow \! \mu^+\nu\gamma$ & $(5.50\pm0.28)\times10^{-3}$ &
KEK-85\cite{kmng}\\
$K^+ \! \rightarrow \! \mu^+\nu\gamma(DE)$  & $(1.33\pm.12\pm.18)\times10^{-5}$
& E787-97\cite{kmng_de}\\ \hline
$K^+ \! \rightarrow \!  e^+\nu\gamma(DE)$ & $(1.52\pm0.23)\times10^{-5}$ &
CERN-79\cite{keng}\\ \hline
$K^+ \! \rightarrow \!  \mu^+\nu\mu^+\mu^-$  & $< 4.1\times10^{-7}$ &
E787-89\cite{kmnmm}\\ \hline
$K^+ \! \rightarrow \!  e^+\nu\mu^+\mu^-$ & $< 5.0\times10^{-7}$ &
E787-98\cite{kenmm}\\ \hline
$K^+ \! \rightarrow \!  \mu^+\nu e^+e^-$  & $\sim$1500 events &
E865-99\cite{kmnee}\\ \hline
$K^+ \! \rightarrow \!  e^+\nu e^+e^-$ & $\sim$400 events &
E865-99\cite{kmnee}\\ \hline
\end{tabular}
\label{tab_kl2g}
\end{table}

\section{Other Radiative Kaon Decays}

The experimental sensitivity for the other radiative kaon decays
K$_{\pi3\gamma}$, K$_{\ell3\gamma}$ and K$_{\pi4\gamma}$ are such as
to only be sensitive to IB contributions. All of these measurements
are consistent with theoretical predictions.  A summary of the results
is given in Table~\ref{tab_kpppg}.
\begin{table}[ht]
\centering
\caption{ \it Summary of Radiative 3- and 4-body decays. }
\vskip 0.1 in
\begin{tabular}{|l||l|l|l|} \hline
Decay Mode  &  \multicolumn{1}{|c|}{Branching Ratio} & Citation \\ \hline\hline
$K^+ \! \rightarrow \!  \pi^+\pi^+\pi^-\gamma$ & $(1.04\pm0.31)\times10^{-4}$ &
ITEP-89\cite{kpppg}\\ \hline
$K^- \! \rightarrow \!  \pi^-\pi^\circ\pi^\circ\gamma$ &
$(7.5^{+5.5}_{-3.0})\times10^{-6}$ & IHEP-95\cite{kpp0p0g}\\ \hline
$K^+ \! \rightarrow \!  \pi^\circ\mu^+\nu_\mu\gamma$ & $< 6.1 \times10^{-5}$ &
ZGS-73\cite{kpmng}\\ \hline
$K^\circ_L \! \rightarrow \!  \pi\mu\nu_\mu\gamma$ &
$(5.7^{+0.6}_{-0.7})\times10^{-4}$ & NA48-98\cite{klpmng}\\ \hline
$K^+ \! \rightarrow \!  \pi^\circ e^+\nu_e\gamma$ &
$(2.62\pm0.20)\times10^{-4}$ & ITEP-91\cite{kpeng}\\ \hline
$K^\circ_L \! \rightarrow \!  \pi e\nu_e\gamma$ &
$(3.62^{+0.26}_{-0.21})\times10^{-4}$ & NA31-96\cite{klpeng}\\ \hline
$K^+ \! \rightarrow \!  \pi^\circ\pi^\circ e^+\nu_e\gamma$ & $< 5\times10^{-6}$
& ITEP-92\cite{kp0p0eng}\\ \hline
$K^+ \! \rightarrow \!  \pi^+\pi^- e^+\nu_e\gamma$  & no limit & (E865-97?)\\
\hline
\end{tabular}
\label{tab_kpppg}
\end{table}

A couple of modes should be seen for the first time in existing data,
$K^+ \! \rightarrow \!  \pi^\circ\mu^+\nu_\mu\gamma$ (E787) and $K^+
\! \rightarrow \!  \pi^+\pi^- e^+\nu_e\gamma$ (E865). Improvements in
other modes may be possible, particularly at IHEP.

\section{Conclusions}

Several new results are expected from KTeV and NA48, as well as a few
more from E787,E865 and E871. With the turn on of DA$\Phi$NE and KLOE,
which is well equipped for the radiative modes, we can expect another
round of new measurements.  Finally, the next generation of rare kaon
experiments, designed to fully constrain the CKM unitarity triangle,
by measuring the `Golden modes' $K^+ \! \rightarrow \! \pi^+ \nu
\overline{\nu}$ and $K^\circ_L \! \rightarrow \! \pi^\circ \nu
\overline{\nu}$ , are under construction (E391a, E949) or being
designed (KOPIO, CKM, KAMI). These experiments will provide even more
precise measurements of several radiative modes.

\section*{Acknowledgements}

I would like to thank several people from various experiments for
providing data and discussions for this talk, including: Hong Ma, Mike
Zeller, Bob Tschirhart, John Belz, Lutz Koepke, Bill Molzon, Leonid
Landsberg, Takeshi Komatsubara, Takashi Nakano and Laurie Littenberg.
This work was supported under U.S. Department of Energy contract
\#DE-AC02-98CH10886.


\begin{thebibliography}{99}
\bibitem{ichep98} Proc. XXIX International Conference on High Energy Physics
(ed. J.~Thomson, TRIUMF, July 1998) (World Scientific,  1999).
\bibitem{dpf99} Proc. 1999 Meeting of DPF
(ed. K.~Arisaka and Z.~Bern, UCLA, Jan. 1999)
http://www.dpf99.library.ucla.edu.
\bibitem{moriond99} Proc. Rencontres de Moriond
(ed. C.~Barthelemy, March 1999).
\bibitem{panic99} Proc. XV$^{th}$ Particles and Nuclei International Conference
(G.~F\"{a}ldt {\it et al}, Uppsala, Sweden, June 1999), (World Scientific,
Singapore, 2000).
\bibitem{epshep99} Proc. International Europhysics Conference on High Energy
Physics
(Ed.~K.Huitu, {\em et. al.}, Tampere, Finland, July 1999), (IOP-Publishing,
Bristol, UK).
\bibitem{kaon99} Proc. Chicago Conference on Kaon Physics
(ed. T.~Quinn, Chicago, June 1999), (University of Chicago Press, 2000).
\bibitem{klppg} J.~Belz, {\em et. al.} in Ref.~\cite{panic99}.
\bibitem{k+ppg} T.K.~Komatsubara {\em et. al.}, in Ref.~\cite{kaon99}.
\bibitem{klppee} K.~Senyo, {\em et. al.} in Ref.~\cite{epshep99}.
\bibitem{klppee2} A.~Alavi-Harati, {\em et. al.} Phys. Rev.  Lett. {\bf 84}, 408 (2000), hep-ex/9908020.
\bibitem{klp0p0g} G.D.~Barr, {\em et. al.}, Phys. Lett. {\bf B328}, 528 (1994).
\bibitem{ksppg} E.J.~Ramberg, {\em et. al.}, Phys. Rev. Lett. {\bf 70}, 2525
(1993), FERMILAB-PUB-92-384-E.
\bibitem{ksppg_de} H.~Taureg, {\em et. al.}, Phys. Lett. {\bf B65}, 92 (1976).
\bibitem{chpt_klpgg} G.~Ecker, {\em et. al.}, Phys. Lett. {\bf B189}, 363
(1987);
G.~D'Ambrosio, {\em et. al.}, Nuov. Cim. {\bf 99A}, 153 (1988).
\bibitem{chpt_av} P.~Heiliger, {\em et. al.}, Phys. Rev.  {\bf D47}, 4920
(1993).
\bibitem{klpgg} A.~Alavi-Harati, {\em et. al.}, Phys. Rev. Lett. {\bf 83}, 917
(1999), hep-ex/9902029.
\bibitem{kpgg} P.~Kitching, {\em et. al.}, Phys. Rev. Lett. {\bf 79}, 4079
(1997), hep-ex/9708011.
\bibitem{klpee} T.~Yamanaka, {\em et. al.} in Ref.~\cite{moriond99}.
\bibitem{klpmm} J.~Whitmore, {\em et. al.} in Ref.~\cite{kaon99}, FERMILAB-CONF-99-266-E.
\bibitem{kpee} R.~Appel, {\em et. al.}, Phys. Rev. Lett. {\bf 83}, 4482 (1999), hep-ex/9907045.
\bibitem{kpmm} H.~Ma, {\em et. al.}, hep-ex/9910047.
\bibitem{kspee} G.D.~Barr, {\em et. al.}, Phys. Lett. {\bf B304}, 381 (1993).
\bibitem{klpeeg} S.~Taegar, {\em et. al.} in Ref.~\cite{dpf99}.
\bibitem{kpeeg} D.~Kraus, {\em et. al.} in Ref.~\cite{dpf99}.
\bibitem{ksgg} G.D.~Barr, {\em et. al.}, Phys. Lett. {\bf B351}, 579 (1995).
\bibitem{klgg} H.~Burkhardt, {\em et. al.}, Phys. Lett. {\bf B199}, 139 (1987), CERN-EP/87-146.
\bibitem{klmm} D.~Ambrose, {\em et. al.} in Ref.~\cite{kaon99}, to be published in {\it Phys. Rev. Lett.}.
\bibitem{pdg} C.~Caso, {\em et. al.}, Euro. Phys. Jou. {\bf C3}, 1 (1998),
pdg.lbl.gov.
\bibitem{chpt_klmm} G.~Valencia, Nucl. Phys. {\bf B517}, 339 (1998), hep-ex/9810007;
D'Ambrosio, {\em et. al.},  Phys. Lett. {\bf B423}, 385 (1998),  hep-ph/9708326.
\bibitem{klee} D.~Ambrose, {\em et. al.}, Phys. Rev. Lett. {\bf 81}, 4309
(1998), hep-ex/9810007.
\bibitem{kleeg} V.~Fanti, {\em et. al.}, Phys. Lett. {\bf B458}, 553 (1999), CERN-EP-99-053.
\bibitem{klmmg} M.B.~Spencer, {\em et. al.}, Phys. Rev. Lett. {\bf 74}, 3323
(1995), hep-ex/9504005.
\bibitem{kleeee} T.~Barker, {\em et. al.} in Ref.~\cite{ichep98}.
\bibitem{klmmee} A.~Lath, {\em et. al.} in Ref.~\cite{dpf99}.
\bibitem{ksmm} S.~Gjesdal, {\em et. al.}, Phys. Lett. {\bf B44}, 217 (1973).
\bibitem{ksee} A.~Angelopoulos, {\em et. al.}, Phys. Lett. {\bf B413}, 232
(1997).
\bibitem{kleegg} T.~Yamanaka, {\em et. al.} in Ref.~\cite{moriond99}.
\bibitem{klmmgg} J.~Whitmore, {\em et. al.} in Ref.~\cite{kaon99}; hep-ex/0001005.
\bibitem{kmng} Y.~Akiba, {\em et. al.}, Phys. Rev. {\bf D32}, 2911 (1985).
\bibitem{kmng_de} M.~Convery, {\em et. al.}, Proc. Meeting of DPF96 (Ed.
H.~Heller, {\em et. al.}),
(Singapore, World Scientific, 1998), to be submitted to Phys. Rev. Lett..
\bibitem{keng} J.~Heintze, {\em et. al.}, Nucl. Phys. {\bf B149}, 365 (1979).
\bibitem{kmnmm} M.~Atiya, {\em et. al.}, Phys. Rev. Lett. {\bf 63}, 2177
(1989).
\bibitem{kenmm} S.~Adler, {\em et. al.}, Phys. Rev. {\bf D58}, 012003-1 (1998), hep-ex/9802011.
\bibitem{kmnee} M.~Zeller, {\em et. al.} in Ref.~\cite{kaon99}.
\bibitem{kpppg} V.V.~Barmin, {\em et. al.}, SJNP {\bf 50}, 421 (1989).
\bibitem{kpp0p0g} V.N.~Bolotov, {\em et. al.}, JETPL {\bf 42}, 481 (1995).
\bibitem{kpmng} D.~Jjung, {\em et. al.}, Phys. Rev. {\bf D8}, 1307 (1973).
\bibitem{klpmng} M.~Bender, {\em et. al.}, Phys. Lett. {\bf B418}, 411 (1998).
\bibitem{kpeng} V.V.~Barmin, {\em et. al.}, SJNP {\bf 53}, 606 (1991).
\bibitem{klpeng} F.~Leber, {\em et. al.}, Phys. Lett. {\bf B369}, 69 (1996).
\bibitem{kp0p0eng} V.V.~Barmin, {\em et. al.}, SJNP {\bf 55}, 547 (1992).
\end{thebibliography}
\end{document}